\documentclass[sn-nature]{sn-jnl}% Style for submissions to Nature Portfolio journals
\usepackage{graphicx}%
\usepackage{multirow}%
\usepackage{amsmath,amssymb,amsfonts}%
\usepackage{amsthm}%
\usepackage{mathrsfs}%
\usepackage{comment}%
\usepackage[title]{appendix}%
\usepackage{xcolor}%
\usepackage{textcomp}%
\usepackage{manyfoot}%
\usepackage{booktabs}%
\usepackage{algorithm}%
\usepackage{algorithmicx}%
\usepackage{algpseudocode}%
\usepackage{listings}%
\usepackage{longtable}%
\usepackage{tabularx}%
\usepackage{adjustbox}%
\usepackage{float}%
\usepackage{lmodern}%
\usepackage{mathrsfs}%

\graphicspath{ {./images/} }
\theoremstyle{thmstyleone}%
\theoremstyle{thmstyletwo}%

\theoremstyle{thmstylethree}%

\begin{document}

%\title[Article Title]{3CFormer, a multitask framework for a contact-aware DNA foundation model}
\title[Article Title]{BMFM-DNA: A SNP-aware DNA foundation model to capture variant effects } %3C-BERT: Pre-training Genomic Foundation Models with Chromatin Interactions}


\author{
\centering
Hongyang Li$^{1,\ast}$, 
Sanjoy Dey$^{1,\ast}$,
Bum Chul Kwon$^{1}$,\\
Michael Danziger$^{2}$,
Michal Rosen-Tzvi$^{2}$,
Jianying Hu$^{1}$,
James Kozloski$^{1}$,
Ching-Huei Tsou$^{1}$, 
Bharath Dandala$^{1,\dagger}$, 
Pablo Meyer$^{1,\dagger}$ \\
\normalsize{$^{1}$IBM Research, Yorktown Heights, NY, USA} \\
\normalsize{$^{2}$IBM Research, Haifa, Israel} \\
\normalsize{$^\ast$These authors contributed equally} \\
\normalsize{$^\dagger$To whom correspondence should be addressed: \\bdand@us.ibm.com, pmeyerr@us.ibm.com}
}

\abstract{Large language models (LLMs) trained on text demonstrated remarkable results on natural language processing (NLP) tasks. These models have been adapted to decipher the language of DNA, where sequences of nucleotides act as ``words" that encode genomic functions. However, the genome differs fundamentally from natural language, as it lacks clearly defined words or a consistent grammar. Although DNA language models (DNALMs) such as DNABERT, GENA-LM have achieved high level of performance on genome-related biological tasks, these models do not encode biological functions in the presence of sequence variations. To address this problem, we pre-train foundation models that effectively integrate sequence variations, in particular Single Nucleotide Polymorphisms (SNPs), as they underlie important biological functions. Specifically, we use ModernBERT to pre-train two different Biomedical Foundation Models (BMFM), namely, BMFM-DNA-REF in which the model is trained with sequences of varying lengths along with their reverse complements derived from the reference genome and BMFM-DNA-SNP in which the model is trained with sequences created using a novel representation scheme that encodes sequence variations. Our findings indicate that integrating sequence variations into DNALMs helps capture the biological functions as seen in improvements on all fine-tuning tasks. To explore the model's practical utility, we experimented with various strategies for SNP imputation on promoter detection task introduced in DNABERT-2. However, we acknowledge that the current benchmarks are limited in their ability to fully evaluate these models. To enable more comprehensive assessment in the future and encourage community contributions, we release our models through HuggingFace and the code to reproduce the results at \href{https://github.com/BiomedSciAI/biomed-multi-omic}{https://github.com/BiomedSciAI/biomed-multi-omic}.}

\maketitle
\section{Introduction}

The advent of high-throughput DNA sequencing technologies has resulted in an unprecedented volume of genomic data, including over 500,000 human genomes, 3,000 genomes from higher organisms, and millions of bacterial genomes. Despite this wealth of data, our understanding of functional genomics remain limited. While the genetic code and non-coding regulatory regions have been well-studied, they account for only about 1.5\% \cite{international2001} and 5–20\% \cite{encode2012}, respectively, of the approximately 3 billion base pairs in the human genome. To address the complexity of interpreting this vast sequence space, machine learning has emerged as a powerful tool. It has been successfully applied to tasks such as gene and splice site identification \cite{jaganathan2019}, prediction of single nucleotide polymorphism (SNP) locations \cite{poplin2018}, functional effect prediction of variants \cite{zhou2015}, and the discovery of regulatory DNA motifs relevant to gene expression \cite{avsec2021}.

Recently, machine learning has experienced a paradigm shift with the rise of general-purpose foundation models, which are trained using self-supervised techniques on large-scale, unlabeled datasets and can be adapted to a wide range of downstream tasks~\cite{bommasani_opportunities_2022}. These models demonstrated impressive capabilities in solving a wide range of problems in natural language processing (NLP) and natural language understanding (NLU) \cite{zhou_comprehensive_2024}. This shift has unlocked new opportunities to study genome function at scale by using models to learn generic  representations from genomic data that can be transferred to quantitative biological tasks. Inspired by this paradigm shift, a growing number of foundation methods have recently been proposed to ``decode the biological language'' by leveraging large-scale genomics, proteomics, transcriptomics, and epigenomics data \cite{si_foundation_2024}. 

Genomic foundation models, often based on transformer architectures \cite{vaswani2017}, extend the central dogma metaphor by treating DNA as a language, with its own words, syntax, and grammar that encode the rules of biology. However, decoding the biological language remains a profound challenge, as the genome, unlike natural language, lacks a clearly defined vocabulary or consistent grammar. It is more akin to a vast manuscript compiled over evolutionary time: layered, fragmented, and overwritten, which some have called a biological palimpsest \cite{haig2004}. In this view, the role of a DNA language model is not just to read a coherent narrative, but to reconstruct meaning from overlapping, context-dependent signals—fragments of many narratives that span millions of years. Another important aspect of genomic variation lies in single nucleotide polymorphisms (SNPs), which are single-base changes in DNA sequence that are present in a large human population. Indeed, SNPs are the most common type of genetic variation among individuals and can influence everything from physical traits to disease susceptibility and drug response \cite{maurano2012}. Understanding the functional impact of SNPs, especially those located in non-coding regions, is crucial for linking genetic variation to phenotypic outcomes and advancing personalized medicine. SNPs in DNA can be compared to polysemous words or tokens in a sentence. Just as each word or token in a sentence carries specific meaning and influences the overall meaning of a sentence, each SNP in a genome represents a small but meaningful variation that can impact biological processes and phenotypic traits.

Genomic foundation models such as DNABERT-2 and GENA-LM \cite{zhou2023dnabert,fishman2025} trained on the human genome and genomes from several species, NT-transformer\cite{dalla2024nucleotide} trained on 3,202 human genomes and 850 genomes from diverse species, and Evo \cite{nguyen2024sequence} trained on 2.7 million prokaryotic and phage genomes have demonstrated success in predicting several biological properties encoded in DNA such as recognizing transcription factor binding sites, predicting gene expression, or predicting promoter activity. 
DNABERT-2\cite{zhou2023dnabert} provided conceptual and empirical insights into genome tokenization, and demonstrated Byte Pair Encoding (BPE), a statistics-based data compression algorithm that constructs tokens by iteratively merging the most frequent co-occurring genome segment in the corpus can encode the genomic language better than k-mer or fixed-length permutations of A, T, C, and G. Other approaches, such as GROVER, have focused on the biological interpretation of BPE tokens \cite{sanabria2024}, demonstrating that token frequencies can reveal insights into genomic patterns. For example, they can highlight a preference for A/T-rich sequences in gene coding regions, which are less prone to mutation, or show a higher GC content near \textit{Alu }regions. However, these models still require additional pre-training tasks to better capture the broader biological context of genomic sequences. Conversely, SEI \cite{chen2022} has adopted a multi-task learning framework to enhance DNA representation learning by combining both self-supervised and supervised objectives. The self-supervised learning objective captures intrinsic relationships between DNA sequences, while the supervised objective leverages available metadata, such as cell type specific DNA accessibility, methylation state, chromatin structure, histone modifications, and external knowledge (e.g. signaling pathway ontology relationships). This dual approach enables the model to encode explicit relationships, facilitating the capture of higher-level biological connections. However, in eukaryotes, questions remain about whether the pre-training strategies of DNA language models (DNALMs) effectively capture key biological properties and consistently outperform traditional approaches \cite{tang2024evaluating, patel2024dart}. In particular, the focus of current DNALMs has been modeling the reference genome, while ignoring the encoding of natural genomic variations in the pre-training stage. This is an important omission, because many drivers of complex inherited disease risk lie in regulatory regions of the genome and SNPs occur almost once in every 1,000 nucleotides on average \cite{maurano2012}. 

To address the above mentioned limitations and further advance this field, we present \texttt{bmfm-multi-omic}, a software package for pre-training, finetuning and benchmarking genomic foundation models. The package supports multiple strategies to encode natural genomic variations; multiple architectures such as BERT, Performer, ModernBERT to build genomic foundation models; fine-tuning and benchmarking of the foundation models on well-established biologically meaningful tasks. In particular, the package incorporates most of the benchmarking datasets from Genomic Understanding and Evaluation (GUE) package released in DNABERT-2. In addition, the package also supports promoter activity prediction on datasets created using Massive Parallel Reporting Assays (MPRA) \cite{agarwal2025}, and SNP-disease association prediction. 

As we continue our journey to expand the software package with additional capabilities, including epigenomic data integration and novel modeling strategies, we invite the research community to collaborate in exploring optimal representations, modeling strategies and creating benchmark datasets for building and evaluating state-of-the-art genomic foundation models.

\section{Datasets}

\begin{figure}[h]
    \centering
    \includegraphics[width=1.0\linewidth]{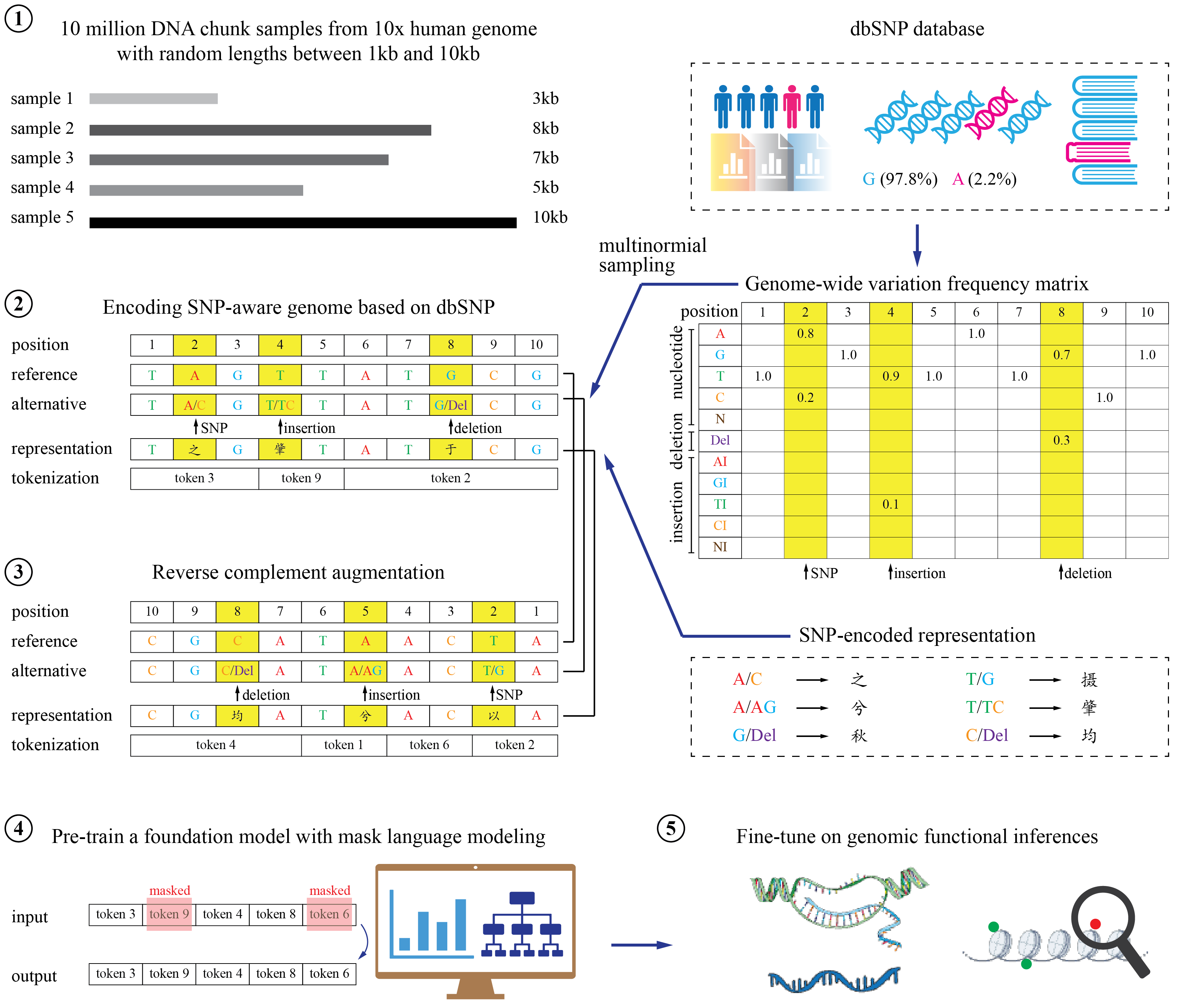}
    \caption{\textbf{BMFM-DNA} The Single Nucleotide Polymorphism Database (dbSNP) is mapped to the human reference genome sequence GRCh38. The genome is sampled 10 times to build DNA sequences of varying length between 1kb to 10kb. Genomic positions with variants are encoded by special characters, where the probabilities of variants are derived from dbSNP. The same process is performed for the reverse complement sequences to enrich the pre-training samples. Sequences are tokenized using BPE algorithm and the pre-training is done with a Masked Language Model (MLM) approach based on a ModernBERT architecture. Finally, model is fine-tuned on several biological tasks including promoter prediction, splicing prediction, transcription factor binding sites prediction, promoter activity prediction using Massive Parallel Reporting Assays (lenti-MPRA), and gene disease associations.}
    \label{fig:Pipeline}
\end{figure}

\subsection{Datasets for pre-training}
\subsubsection{Human Genome}
We used the genome assembly GRCh38 (Genome Reference Consortium Human Build 38) \cite{schneider2017} downloaded from https://hgdownload.soe.ucsc.edu/downloads.html 

\subsubsection{dbSNP}
dbSNP is a public repository of submitted nucleotide variations integrating data from large-scale sequencing projects such as 1000 Genomes, gnomAD, TOPMed, and ALFA. It currently hosts 5.0 billion submitted SNPs and 1.2 billion unique reference SNPs, making it one of the most comprehensive resources for human genetic variation. \cite{smigielski2000}.

\subsection{Datasets for finetuning}
We incorporate finetuning datasets from diverse sources for each of the following tasks that are biologically meaningful as described below.

\subsubsection{Promoter detection} The \textit{promoter detection} is a binary classification task that focuses on identifying proximal promoter region sequences in the human genome. The dataset \cite{zhou2023dnabert, fishman2025} consists of two types of promoters: those containing the TATA box motif and those that lack the TATA box motif. The -249 to +50 bp region surrounding the transcription start site (TSS) was extracted from both TATA and non-TATA promoters obtained from the Eukaryotic Promoter Database (EPDnew) \cite{dreos2013epdnew}, and used to define the promoter class. The non-promoter class was created using two approaches: 1) randomly sampling sequences outside of the promoter regions that contain the TATA motif, and 2) randomly shuffling consecutive, non-overlapping 20-bp subsequences from non-TATA promoter sequences (non-TATA, non-promoters). This method for generating negative samples was adapted from \cite{fishman2025, oubounyt2019deepromoter}. The total number of sequences is N=47,356, with N=5,920 for training, N=5,920 for development, and N=5,920 for testing.

%either by taking a random set of Chinese characters from the 121 representing biallelic variants (Class 1), matching the locations of the SNPs in the positive set to the negative set ((Class 2), or by taking the Chinese characters representing the bi-allelic SNPs in the positive set and randomizing their locations in the negative set (Class 3). Finally we created a new set of negative promoters by just taking the positive class and randomizing the SNP positions (Class 4). 

\subsubsection{Core promoter detection} The \textit{core promoter detection} is a binary classification task and is similar to promoter detection with a focus on predicting the core promoter region only, the central region closest to the TSS and start codon. A much shorter context window (centered -34 +35 bp around TSS) is provided, making this a more challenging task than promoter prediction. The total number of sequences are (N=47,356) train, (N=5,920) dev, (N=5,920) test.

\subsubsection{Transcription factor binding site prediction} The \textit{Transcription factor binding site prediction} is a classification task extracted from \cite{zhou2023dnabert} aiming to predict sequences that contain binding sites for transcription factors (TFs). These sites can be activator or repressors of gene expression, typically they are 5 to 20 base pairs in length and are often found in the promoter region, but can also exist elsewhere in the genome. The legacy 690 ENCODE ChIP-seq experiments \cite{encode2012}, available through the UCSC Genome Browser, include 161 TF binding profiles across 91 human cell lines, from which a 101-bp region was extracted around the center of each peak to represent the TF binding site (TFBS) class, with non-overlapping sequences of the same length and GC content forming the non-TFBS class. Finally, 5 datasets were randomly selected out of a subset of 690 curated by heuristically filtering out tasks that are either too trivial (e.g., over 0.95 $F_1$) or too challenging (e.g., less than 0.50 $F_1$) for models. The total sequences in five training sets are 32,378, 30,672, 19,000, 27,294, 19,000, while dev and test set have 1,000 samples in each dataset.

\subsubsection{Splice site prediction} The \textit{splice site prediction} is a classification task aiming to predict whether a sequence contains a splice donor or a splice acceptor site or neither. We constructed the dataset from \cite{wang2019}, which consists of 10,000 400-bp long sequences. Additionally, we added the adversarial examples to make this task more challenging introduced in \cite{zhou2023dnabert}. The total sequences are (N=36,492) train, (N=4,562) dev, (N=4,562) test.

%\subsubsection{Prediction of chromatin profiles} Predicting the epigenetic states of a locus from its sequence is another pervasive challenge in genomics. We utilized the DeepSEA dataset \cite{zhou2015deepsea} as done in \cite{fishman2025} to evaluate the capacity of BMFM-DNA to tackle this challenge. This dataset contains 919 cell-type specific chromatin profiles, categorized as DNAse I hypersensitivity sites (DHS) that can indicate nucleosome-free region harboring regulatory factors that alter chromatin structure, histone marks (HM) also producing changes to histone proteins that package DNA into nucleosomes, and transcription factor binding sites (TF). In the original DeepSEA challenge, the signals of these chromatin marks were predicted for each 200 bp genomic bin, using its sequence and the sequence of the 800 bp context (±400 bp flanking regions). The task was setup as a multiclass-multilabel problem with \textit{BCEwithLogistic} loss, where DHS, HM and TF are set as three classes and each class can have multiple labels (125, 104 and 690 labels for three classes respectively). The total sequences are (N=4400000) train, (N=8000) dev, (N=455024) test. 

\subsubsection{Massive Parallel Reporting Assay}
To further validate the generalizability of our models, we used lentivirus-based MPRAs (lenti-MPRAs) dataset to measure the gene expression induced by Cis Regulatory Elements (CREs) across three human cell types for lymphoblasts (K562). MPRA tests thousands of sequences or variants for regulatory activity in a multiplex fashion and provides a readout as the virus randomly integrates in genome loci, it can test more than 200,000 sequences in a single experiment. For the large-scale K562 lenti-MPRA library, every sequence and its corresponding reverse complement were grouped together and these pairs were distributed into train (N=143295), dev (N=39,370), and test (N=43,588) to ensure that both the forward and the reverse sequences resided within the same dataset. The human MPRA dataset is available from Zenodo (https://doi.org/10.5281/zenodo.8219231).

\subsubsection{Predicting the association between SNPs and diseases}
Disease and trait-associated genetic variants have been identified with genome-wide association studies (GWAS), and there is a vast literature building models that predict gene associations with disease \cite{pinero2020disgenet}. However, the majority (~93\%) of disease and trait-associated variants lie within non-coding regions of the genome. Hence, we decided to build a model that goes further than coding regions and can also predict links between target diseases and SNPs independently of their location~\cite{maurano2012}. 
To perform this task, we fine-tuned the BMFM-DNA models to estimate the probability of an SNP being associated with a disease: $\mathrm{P}(S|D)$. We used GWAS Catalog\cite{cerezo2025} and ClinVar\cite{landrum2019clinvar} to prepare the dataset. We combined the individual associations between each SNP and disease from ClinVar and GWAS Catalog (N=3,004,727). Then, we excluded associations with no or little reported significance (e.g., association not found) and merged the disease ontologies (i.e., EFO GWAS Catalog \cite{malone2010} and MONDO \cite{vasilevsky2022} for ClinVar) into one (N=1,381,365). For each SNP, we extracted 1,001 nucleotides, including the SNP location, along with 500 base pairs upstream and 500 base pairs downstream. In total, we had 475,694 unique SNPs extracted as DNA sequences.
We used the top 2,000 out of 14,435 most frequently occurring diseases, which range from fronto-temporal dementia (N=72), inborn genetic disease (N=18,936), hereditary neoplastic syndrome (N=61,000), to non-specified (N=429,088).
We split the 475,694 sequences into the ratio of 0.8, 0.1, and 0.1 for train (N=380,555), dev (N=47,569), and test (N=47,570) sets, respectively.

\subsection{Variant-encoded datasets} \label{sec:finetune_datasets}
For the promoter detection dataset mentioned above, we created a variant-encoded version of input DNA sequences through a two-step process. First, if the location of the sequence was provided in the dataset, we used them; otherwise, we aligned the sequence to the reference genome via the standard sequence alignment tool, \textit{bwa-mem} \cite{li2013aligning}.
Next, we identified variations within the corresponding genomic region based on dbSNP, and then introduced variants into the sequence using the sampling algorithm outlined in \ref{bmfm-dna-snp}. However, in some datasets, particularly those where negative samples were artificially created, alignment was challenging and also not applicable. Thus, we used four different techniques to create variant-encoded negative samples as described below:

\begin{itemize}
     \item \texttt{Class 1}: We used \textit{bwa-mem} alignment tool for both positive and negative sets. 
     \item \texttt{Class 2}: We calculated the frequency distribution of SNPs versus non-SNPs in the positive sequences. Then, we randomly sampled from the 121 possible SNPs to impute the negative sequences at random locations, while respecting the frequency distribution.
    \item \texttt{Class 3}: We calculated the frequency distribution of SNPs in the positive set. Then, based on this distribution, we imputed SNPs at random locations within each negative sequence.
    \item \texttt{Class 4}: We used the positive samples and inserted SNPs at random locations, treating these modified samples as negative examples.
\end{itemize}

\section{Methods}
DNALMs are based on the powerful transformer architecture \cite{vaswani2017attention}, which excels at capturing long-range dependencies within data, making it well-suited for analyzing complex relationships between DNA sequences. As DNA encodes information in a bi-directional multi-strand way, we used ModernBERT \cite{warner2024}, a modern bidirectional encoder for fast, memory efficient, and long context training and inference, bringing a substantial improvement over older encoders. We pre-trained two models using ModernBERT architecture with the masked language modeling objective. The schematic diagram of our pipeline is shown in Figure \ref{fig:Pipeline}.

\subsection{Data preparation for pre-training}
\subsubsection{\texttt{BMFM-DNA-REF}}
The pre-training samples were prepared by extracting DNA sequences of random lengths (between 1kb and 10kb) consecutively from the human reference genome. Sequences were excluded if all nucleotides are “N”. To further enrich the diversity of the training set, we repeated the whole-genome random sampling 10 times. For each DNA sequence sample, we also created the reverse complement sequence as the counterpart, leading to a total of 9,982,678 samples that roughly cover the human genome 20 times or about 60 billion nucleotides.
\subsubsection{\texttt{BMFM-DNA-SNP}}
\label{bmfm-dna-snp}
In order to take advantage of the biological information encoded in SNPs, we first extracted 20 million genetic variants from Single Nucleotide Polymorphism Database (dbSNP), including SNPs, insertions and deletions. For each genomic position that has variants, we averaged available population-specific variation frequencies and built a genome-wide variation frequency matrix. For each genomic position, this matrix provides the multinomial distribution of 11 probabilities on 5 types of nucleotides (A, C, G, T, N), 5 types of insertions (insertions after A, C, G, T, N), and deletion.  For genomic positions that do not have alternatives, we assign the probability of 1.0 to the nucleotide based on the human reference genome GRCh38. Through this genome-wide variation frequency matrix, we can efficiently retrieve biallelic variants based on population frequencies and sample variant-encoded DNA sequences. For more details, please refer to \cite{cahyawijaya2022}.
Next, similar to the reference genome, we created random DNA sequence samples from the variation-encoded genome. Specifically, for each genomic position with variations, we first sampled a biallelic representation of two possible nucleotides, insertions, or deletions without replacement from the multinomial distribution of the variation frequency matrix (see Figure \ref{fig:Pipeline}). Then we mapped each variant or nucleotide-pair of variants to a unique single Chinese character for a total of 121 characters, all extracted from the classic poem \textit{Li Sao}. In this way, the nucleotide variations are encoded into the pre-training samples, whereas two possible nucleotides at a single position are still represented by one character instead of two characters.

%We extracted sequences of different size from the human reference genome GRCh38 \cite{schneider2017} where, most of the positions are mapped and represented as either ‘A’, ‘T’, ‘C’, or ‘G’, while the others are unmapped and flagged with the unknown (‘N’) token. We use dbSNP to find the positions that encode SNPs in the reference genome, as it is a central public repository of human SNPs that covers a broad collection of simple genetic variations with a length of variation over 50 base pairs long, including single-base nucleotide substitutions, small-scale multi-base deletions, and small-scale multi-base insertions \cite{smigielski2000}. As additional sources of information from other organisms or different architectures make harder to follow what the model is learning and to link the learned representations back to the relevant genome sequence. Therefore, we decided to train only with the human genome sequence, distributed into tokens.

%\subsubsection{Reference and variant-encoded pre-training DNA sequence samples}
%In this model, we combined the reference genome and variation-encoded genome together. We followed the similar sampling strategy to create sequences. 

\subsection{Tokenization of pre-training DNA sequence samples}
DNALMs can be broadly categorized into four types based on their input representation: one-hot encoded vectors, nucleotide k-mers, single nucleotide tokens, or tokenized sequences using algorithms like Byte-Pair Encoding (BPE). Recent research\cite{sanabria2024, sennrich2015} has highlighted the benefits of using BPE algorithm to construct a vocabulary of tokens representing k-mers of varying lengths, based on their frequency in the genome, rather than relying on fixed-length k-mers. Inspired by this research, we created a tokenizer using BPE algorithm with a vocabulary size of 4,096 tokens for all three approaches. It is worth noting that the tokenizers’ vocabularies differ between the two approaches. We trained SentencePieceBPE tokenizer \cite{kudo2018} using 20kb DNA sequences, which were longer than the pre-training samples. Two tokenizers were trained one with the reference, and other with the variation-encoded DNA sequences, respectively. Both tokenizers had a vocabulary size of 4096 tokens, balancing the performance and computational efficiency. To accelerate data processing and model training, the reference and variation-encoded samples were pre-tokenized using the corresponding tokenizers and converted into LitData \cite{litdata2023} format before pre-training. Note that only part of the 121 Chinese characters are used in the variant-encoded tokenizer vocabulary, since some of the combinations do not exist in the dbSNP together with the human genome.

\subsubsection{Comparative analysis of tokenizers}
 
\begin{figure}[h]
    \centering
    \includegraphics[width=0.9\linewidth]{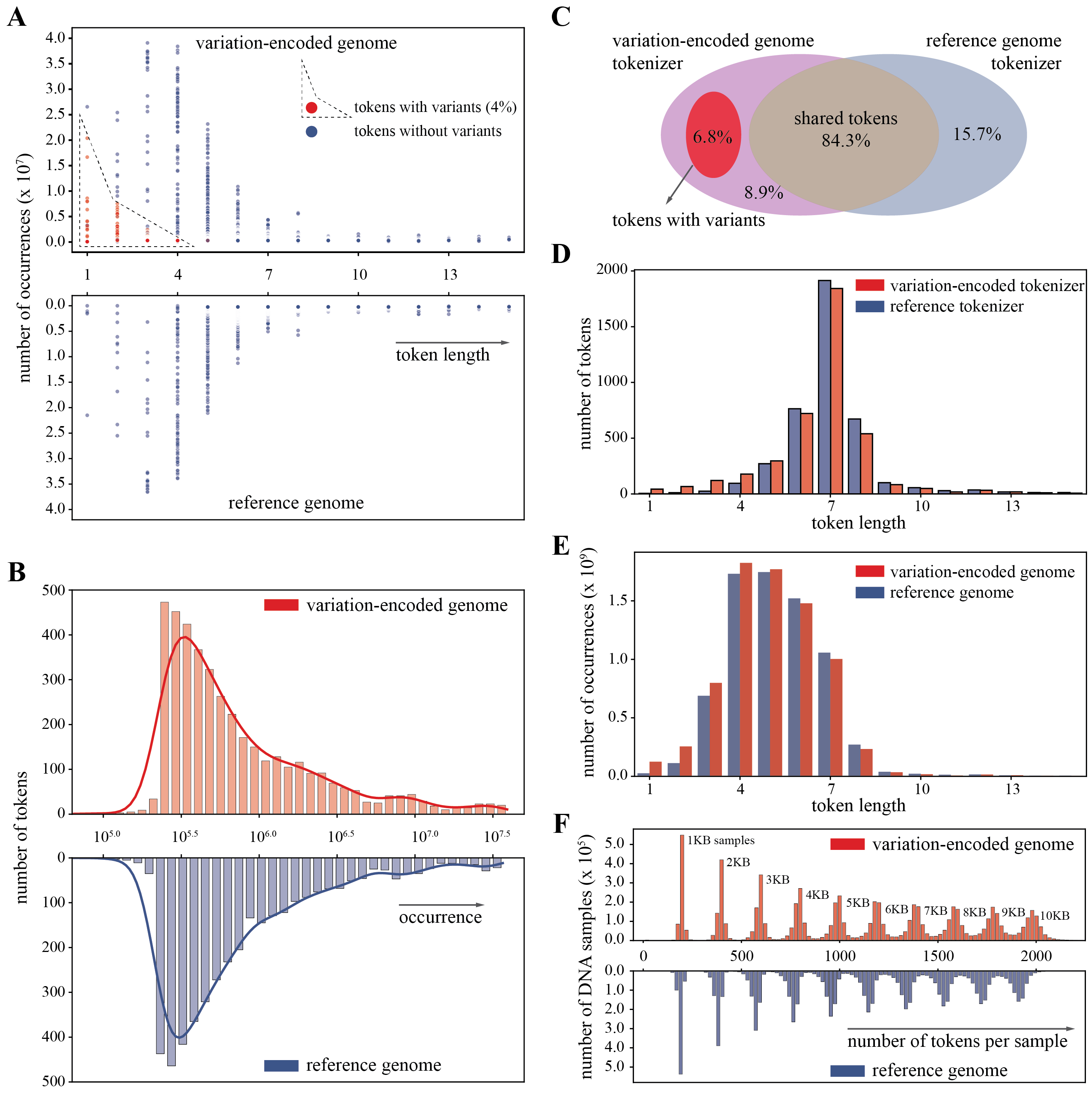}
    \caption{\textbf{Comparative analysis of genome tokenization.} To learn the effects of genetic variations in the human genome, we built two tokenizers for the variation-encoded and reference genomes, respectively. The two versions of the genome are tokenized and the distributions of genome-wide tokens are presented as scatter plots \textbf{(A)} and histogram plots \textbf{(B)}. The vocabularies of two tokenizers are compared through a Venn diagram \textbf{(C)}, where most tokens are shared and the proportion of tokens with variants is highlighted. We further compared token length distributions from the vocabularies \textbf{(D)} and whole genomes \textbf{(E)}. The number of tokens per DNA sample are compared as well \textbf{(F)}, 10 peaks emerge corresponding to their nucleotide lengths ranging from 1kb to 10kb.}
    \label{fig:tokens}
\end{figure}

 We performed a comparative analysis of the genome-wide tokenization of the reference and variant-encoded genomes (Figure \ref{fig:tokens}). As shown by the scatter plot in Figure \ref{fig:tokens}A, each dot represents one of the 4091 tokens (5 special tokens were excluded) from the variation-encoded tokenizer (top) and the reference tokenizer (bottom), special tokens are not accounted. Overall, the size distributions of the tokens is similar for the two versions of the human genome, and most tokens are shorter than 10 nucleotides. Tokens with variants (red) are even shorter and are mainly distributed to the left, accounting for 4\% of all tokens in the variation-encoded genome. We next analyzed the histogram distributions of token occurrence in both genomes (Figure \ref{fig:tokens}B), where most tokens occurred more than 300,000 times (around $10^{5.5}$) and the overall distributions are also similar. 

Intriguingly, vocabularies of the variation-encoded and reference tokenizers have 84.3\% of the tokens in common (Figure \ref{fig:tokens}C). Within the variation-encoded tokenizer, 6.8\% of tokens have variants, much higher than the natural abundance of genomic variations which is around 1\%. This indicates that the observed genomic variations have specific patterns of repetition in the human genome and these were learned during BPE tokenizer training, leading to a relatively higher proportion of tokens with variants. Analysis of the token length distribution for the two versions of the genome reveals that the most abundant token length in the two vocabularies is of 7 nucleotides (Figure \ref{fig:tokens}D) with more than 1,500 tokens in both variation-encoded and reference tokenizers. Compared to the reference tokenizer, the variation-encoded tokenizer (red) has shorter tokens in their vocabulary as we expected given the necessity to represent existing variants. Analysis of the token length distribution in the genome instead of the vocabulary (Figure \ref{fig:tokens}E), reveals that the overall distribution is shifted towards smaller token lengths. In contrast to the single high peak at the token length of 7 nucleotides in the vocabulary, token lengths of 4 and 5 nucleotides are the most abundant. This illustrates that for both versions, shorter tokens occur more frequently in the genome than in the model's vocabulary. Finally, we calculated the number of tokens for each DNA sample. The number of tokens per sample in the whole genome is shown as two histogram plots in Figure \ref{fig:tokens}F. Our implementation of randomly creating DNA sequences of different lengths ranging from 1kb to 10kb, is reflected in the observed 10 peaks in both genome versions. In general, samples from the variation-encoded genomes have more tokens, consistent with the fact that tokens from the variation-encoded genome are shorter. The maximum number of tokens per sample being around 2,000, we decided to set the maximum number of tokens to 2,048 in our model.

\begin{figure}[h]
    \centering
    \includegraphics[width=0.9\linewidth]{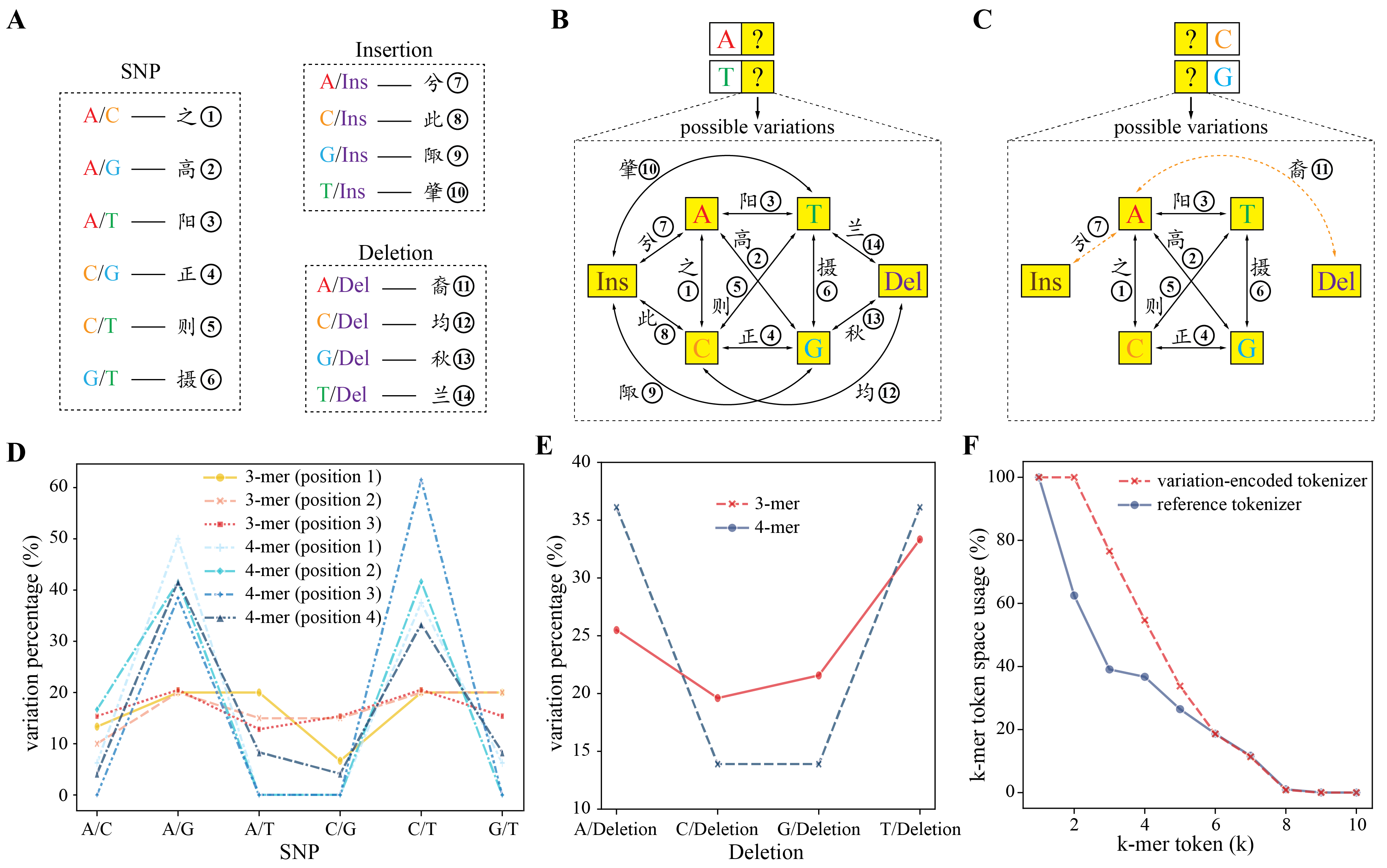}
    \caption{\textbf{Variant patterns in the variation-encoded tokenizer vocabulary.} \textbf{(A)} We encode variations in the human genome by mapping each variant (SNP, Insertion or Deletion) to a unique Chinese character. \textbf{(B)} The observed variants from 2-mer tokens starting with A or T are represented as a graph. Vertices of the graph represent four nucleotides, insertion, and deletion, whereas edges represent variants. If an edge exists in the graph, the corresponding variant is observed in 2-mer tokens of the variation-encoded tokenizer vocabulary. \textbf{(C)} The observed variants from 2-mer tokens ending with C or G are represented as a graph. The orange dashed edges represent variants only observed in tokens ending with \textbf{C}. For 3-mer and 4-mer tokens, the major components are single-variant tokens with a single SNP and all-variant tokens with multiple deletions. The position-specific percentages of six SNPs in single-variant tokens are shown in \textbf{(D)}. The overall percentage of four deletion variants in all-variant tokens are shown in \textbf{(E)}. The k-mer token space usage in the reference and variation-encoded tokenizer vocabularies are shown in \textbf{(F)}. The k-mer token space usage is defined as the number of observed k-mers divided by the number of all possible k-mers of four nucleotides. For the variation-encoded tokenizer, tokens with variants are mapped to the non-variant reference tokens before calculating the k-mer space usage.}
    \label{fig:tokensa}
\end{figure}

We further analyzed short k-mer tokens with characters representing nucleotide variants. The mapping between genomic variants and special Chinese characters is shown in Figure \ref{fig:tokensa}A. Specifically, two types of 2-mers dominate the vocabulary: (1) tokens beginning with nucleotide A or T and ending with variants, and (2) tokens beginning with variants and ending with nucleotide C or G (Figure \ref{fig:tokensa}B-C). The possible variants can be represented as edges in a graph of A, C, G, T, insertion (Ins), and deletion (Del). For example, a SNP A/T is represented as the edge between vertices A and T. In the first type of tokens beginning with A or T, most variations exist in the vocabulary, except for A/deletion and insertion/deletion (Figure \ref{fig:tokensa}B). Whereas in the second type of tokens ending with C or G, all SNPs exist, and most insertions and deletions are not present (Figure \ref{fig:tokensa}C). The only exceptions are A/insertion – C and A/deletion – C. Moreover, we analyzed 3-mers and 4-mers with a single variant or multiple variants. The variants are always SNPs in single-variant tokens, while the variants are deletions in all-variant tokens. Regarding the single-variant tokens, we calculated the percentage of different variations at each position in 3-mers and 4-mers (Figure \ref{fig:tokensa}D). In 3-mers, the percentage distributions are relatively even, and each SNP has a similar presence around 10\%-20\%. However, in 4-mers, two SNPs (A/G and C/T) are dominating with 40\% of occurrences each, and the other four types of SNPs only have less than 10\% of occurrences. Similar results are observed for the all-variant tokens. Four deletions are relatively equally distributed in 3-mers, whereas A/deletion and T/deletion have more than 35\% of occurrences in 4-mers (Figure \ref{fig:tokensa}E).

In summary, these results indicate that hidden patterns of variation distributions can be revealed through building tokenizers on the variation-encoded genome. Although these tokens with variants only contribute to 6.8\% of the variation-encoded tokenizer (Figure \ref{fig:tokens}D), each token represents combinations of nucleotides and multiple possibilities of DNA sequences, increasing the pre-training DNA sample space. For example, a SNP position with the reference nucleotide A and the alternative C is represented by a special Chinese character. Similar to the degenerate state in quantum mechanics, a special character of variation implicitly introduces multiple possible nucleotides, insertion or deletion at one genomic position, where the possibilities are reflected by the variation frequency matrix based on dbSNP. Therefore, the vocabulary in the variation-encoded tokenizer could have higher k-mer token space usage compared to the reference genome tokenizer vocabulary. We focused on the shorter k-mers between 1-mers and 10-mers (Figure \ref{fig:tokensa}F), since they are the major component of the vocabulary. We calculated the k-mer space usage, which is the number of observed k-mers divided by the number of all possible combinations of four nucleotides A, C, G, and T. For the variation-encoded tokenizer, we mapped each token with variants to all possible tokens based on the variation encoding. For example, a token with SNP of T-G-A/C is mapped to two reference tokens of T-G-A and T-G-C. Compared to the reference genome, the variation-encoded vocabulary has a much higher k-mer space usage for 2-mer, 3-mer, 4-mer, and 5-mer (Figure \ref{fig:tokensa}F).

\subsection{ModernBERT Architecture}
ModernBERT is a modernized encoder-only transformer model, extending the standard BERT architecture\cite{devlin2018bert} that incorporates recent advances on improved efficiency and performance, especially for longer sequence lengths. Specifically, ModernBERT adapts several architectural improvements such as use of Rotary Positional Embeddings (RoPE) instead of absolute embeddings, GeGLU activation (a variant of GLU) instead of GeLU is used for better empirical performance, pre-normalization block with LayerNorm after the embedding layer and additionally removing the first LayerNorm, and disabling bias terms. It also adapts several efficiency improvements such as use of FlashAttention and its unpadding capabilities to improve memory and compute efficiency, and alternative attention in which attention layers alternate between global attention, where every token within a sequence attends to every other token, and local attention, where tokens only attend to each other within a small sliding window. The parameters are summarized in Table \ref{tab:model}.

\begin{table}[htbp]
  \centering
  \caption{\texttt{BMFM-DNA} Model Hyperparameters}
  \label{tab:model}
  \begin{tabular}{|l|c|}
    \hline
    \textbf{Hyperparameters} & \textbf{Value} \\
    \hline
    Num. of attention heads & 12 \\
    Num. of hidden layers & 22 \\
    Hidden size & 768 \\
    Global rope theta & 160,000.0 \\ 
    Local rope theta & 10,000.0 \\
    Local attention  & 128 \\
    Global attention every n layers & 3 \\
    Intermediate size & 1,152 \\
    Vocab size & 4,096 \\
    Max. sequence length & 2,048 \\
    \hline
  \end{tabular}
\end{table}

\subsection{Implementation}
We pre-trained our foundation models using the masked language modeling (MLM) objective. To minimize the discrepancy between pre-training and fine-tuning tasks (where fine-tuning samples typically do not include masked tokens), we followed a strategy where 90\% of the tokens were replaced with [MASK], while the remaining 10\% were left unchanged, similar to the approach used in DNABERT-2. During pre-training, we optimized the MLM objective using cross-entropy loss and monitored perplexity to assess performance.

The 10 million DNA sequence samples were randomly split into training (70\%), validation (20\%), and testing (10\%) sets. The models were pre-trained for 150,000 steps, which is shorter compared to DNABERT-2, pre-trained for 500,000 steps. We use a batch size of 40 and a max sequence length of 2,048. We use the AdamW optimizer with $\beta_1$ =0.9, $\beta_2$ =0.98, $\epsilon$=$10^{-6}$ and weight decay of $10^{-5}$. The learning rate linearly increases from 0 to $5\times10^{-4}$ during the first 30,000 steps, then linearly decreasing in the last 120,000 steps. The pre-training took approximately 10 days using 4 NVIDIA A100-SXM4-80GB GPUs.

\section{Experiments}

\subsection{Baseline}
We compare BMFM-DNA with DNABERT-2 \cite{zhou2023dnabert}, a state-of-the-art genomic foundation model. DNABERT-2 utilizes a BPE tokenizer and incorporates several strategies to address input length constraints, reduce time and memory usage, and enhance model performance.

For all benchmark tasks, DNABERT-2 serves as the baseline except for the SNP-to-disease association prediction. In the SNP-to-disease association prediction task, we compare our results to the deep-learning model EVE \cite{frazer2021EVE}.

%\textbf{GENA-LM} are suite of transformer-based foundation DNA language models capable of handling input lengths up to 36 000 base pairs. It includes multispecies and taxon-specific models, demonstrating their capability for fine-tuning and addressing a spectrum of complex biological tasks with modest computational demands.
%and GENA-LM \cite{fishman2025} depending on the fine-tuning tasks. 
\subsection{Setup and Metric}
During finetuning, we keep the majority of hyperparameters (e.g., learning rate, batch size, weight decay) consistent across all tasks, adjusting only the maximum sequence length and the number of training steps based on the dataset's requirements. We validate the model every 200 steps and save the checkpoint with the lowest validation loss. Final results are reported using the test set. To evaluate model performance, we use $F_1$ score, Matthews Correlation Coefficient (MCC), and area under the receiver operating characteristic curve (AUC). For each model, we train with three different random seeds and report the average performance. The overall performance of each model on specific tasks is summarized in Table \ref{tab:finetuning}. 

%\subsection{Main results: pre-training}

%\begin{figure}[h]
 %   \centering
  %  \includegraphics[width=0.9\linewidth]{images/fig3.png}
   % \caption{\textbf{UPDATE THIS FIGURE Pre-training using MLM task, reference and variation-encoded genomes.}}
    %\label{fig:pre-training-results}
%\end{figure}

We finetuned both \texttt{BMFM-DNA-REF} and \texttt{BMFM-DNA-SNP} on 6 tasks, out of which 4 were introduced in DNABERT-2's GUE, i.e transcription factor, promoter prediction, core promoter and splicing detections \cite{zhou2023dnabert}. The remaining two were lenti-MPRA task introduced in \cite{rafi2024} and derived from \cite{agarwal2025}, and the SNP-to-disease association task developed for this manuscript but is similar to gene-to-disease tasks described in \cite{frazer2021EVE,cheng2023,pinero2020disgenet}. 
\subsection{Experimental Results}

As shown in Table \ref{tab:finetuning}, our results with \texttt{BMFM-DNA-REF} and \texttt{BMFM-DNA-SNP} 
achieved similar level of performance when compared with DNABERT-2. It is worth noting that DNABERT-2 is trained on a much larger dataset, integrating genomes from 135 species. Furthermore, it is more suitable for shorter sequences, such as those used in our current fine-tuning tasks, as it is pre-trained on sequences of length 128 (as shown in \cite{fishman2025}). More importantly, the SNP-aware pre-trained model \texttt{BMFM-DNA-SNP} outperformed the \texttt{BMFM-DNA-REF} on all of the tasks, except for the SNP-to-disease task where no improvement was seen, demonstrating the importance of encoding reference genome along with sequence variations in building DNALMs.

\begin{table}[htbp]
  \centering
  \caption{\textbf{Fine-tuning results of multiple tasks (row) from different models (column)}. 
  \newline Depending on the task type, various scoring metrics are used as labeled in the Task column. We compare DNABERT-2 (EVE for SNP-to-disease) with three versions of our models, including a model without pretraining (BMFM-DNA no-pre-training), a model pre-trained on the reference genome (BMFM-DNA-REF), and a model pre-trained on the variation-encoded genome (BMFM-DNA-SNP).} %Note that the last model has two sets of performance scores. The left score is based on non-variant DNA sequences as input, whereas the right score is based on variation-encoded sequences during fine-tuning. Bold represents our best result when there is one.}
  \begin{tabular}{lcccc}
    \hline
    \textbf{Task} & \parbox{2cm}{\centering\texttt{DNABERT-2/EVE}} & \parbox{2cm}{\centering\texttt{\newline BMFM-DNA no-\newline} \texttt{pre-training}}  & \parbox{2cm}{\centering\textbf{\newline \texttt{BMFM-DNA-REF}} \newline} & \parbox{2cm}{\centering\textbf{\newline \texttt{BMFM-DNA-SNP}} \newline} \\
    \hline
    Promoter ($F_1$)  &  94.00  & 89.67 & 92.08 & 93.5\\
    cPromoter ($F_1$) & 83.85  & 77.73 & 81.97  & 83.19 \\
    TF binding ($F_1$) & 83.91  & 75.19 & 81.87  & 82.42  \\ 
    Splicing ($F_1$) & 90.66  & 68.18 & 89.29 & 90.44 \\ 
    
    \hline
    lenti-MPRA K562 (PCC)  &  75.00  & 65.12 & 74.70 & 75.53 \\
    SNP to Disease (AUC) &   91.00  & 87.00 & 90.00 & 90.00 \\
    \hline
  \end{tabular}%
  \label{tab:finetuning}%
\end{table}%

\begin{table}[htbp]
  \centering
  \caption{\textbf{ Performance of BMFM-SNP on Promoter Detection Task}: Fine-tuning results using different variant-encoded strategies for promoter detection. Each class represents a different method for generating negative samples, with the effectiveness measured by $F_1$ score. For Class 4, we compare the performance of ModernBERT trained from scratch (left) with the performance of a fine-tuned \texttt{BMFM-DNA-SNP} model (right).}
  \begin{tabular}{lccccc}
    \hline
    \textbf{Task} & \parbox{1.5cm}{\centering\textbf{\newline No-variant encoded}\newline}& \parbox{1.5cm}{\centering\textbf{Class 1} \newline (impute SNPs from alignment)} & \parbox{1.5cm}{\centering\textbf{Class 2}\newline (random SNPs)}  & \parbox{1.5cm}{\centering\textbf{Class 3} (randomized same SNPs as + class) } & \parbox{1.5cm}{\centering\textbf{Class 4 } (same sequence randomized SNPs)} \\
    \hline
    Promoter ($F_1$)  &  93.5 &   93.01 &  97.46 & 94.10 & 57.04/64.6 \\
    \hline
    \hline
  \end{tabular}%
  \label{tab:negatives}%
\end{table}%

%\begin{table}[htbp]
%  \centering
%  \caption{\textbf{Performance of BMFM-SNP}. Fine-tuning results for SNPs imputed in all datasets of the multiple tasks (row) and when applicable, different imputation of variants in the negative class (columns). Bold represents our best result when there is one. }
%  \begin{tabular}{lccccc}
%    \hline
%    \textbf{Task} & \parbox{1.5cm}{\centering\textbf{\newline No-variant encoded}\newline}& \parbox{1.5cm}{\centering\textbf{Class 1} \newline (random SNPs)} & \parbox{1.5cm}{\centering\textbf{Class 2}\newline (impute SNPs from alignment)}  & \parbox{1.5cm}{\centering\textbf{Class 3} (randomized same SNPs as + class) } & \parbox{1.5cm}{\centering\textbf{Class 4 } (same sequence randomized SNPs)} \\
 %   \hline
%    Promoter ($F_1$)   &  93.5 &  \textbf{97.46}  & 93.01 & 94.10 & 57.04/64.6 \\
%    cPromoter ($F_1$)  &  83.19 & \textbf{87.01}	& 76.33	& 83.83 & -\\
%    TF binding ($F_1$)  &  82.42 & \textbf{83.02} & - & - & - \\ 
%    Splicing ($F_1$)  &  90.44 & \textbf{90.46 } & - & 89.60 & - \\
%     \hline
%    lenti-MPRA K562 (PCC)  & \textbf{75.53}  & - & 73.82 & - & - \\ 
%    SNP to Disease (AUC)   & 90.00  & - & 90.00 & - & - \\ 
%    \hline
%    \hline
%  \end{tabular}%
%  \label{tab:negatives}%
%\end{table}%

To further evaluate the impact of SNPs, we finetuned the \texttt{BMFM-DNA-SNP} model on the variant-encoded datasets introduced in section \ref{sec:finetune_datasets}. We report the results in Table \ref{tab:negatives}.

%When the variation-encoded datasets were used, both the promoter task and core promoter task performed better surpassing DNABERT-2 by almost 4 points (see Table \ref{tab:finetuning} column 4 right side). The splice task performed better but did not pass DNABERT-2, the lenti-MPRA task did not perform better and the SNP-to-disease task  performance was constant (See Table \ref{tab:finetuning} column 4 left and right side). It is to note that BMFM-SNP beat SOTA for the promoter tasks by a large margin only for class 1 negatives, but also does for class 3 negatives (see Table \ref{tab:negatives}). The performance of the BMFM-SNP model on the promoter prediction task for class 4 negatives, was downgraded to $F_1$=64.6/ MCC=30.29. However, the model without pre-training had a way lower performance of 57.04/17.42 or 13\%/50\% less (see Table \ref{tab:negatives}). This indicates that BMFM-SNP is indeed able to learn the DNA sequence context where SNPs appear and use it to distinguish two identical sequences of promoters where the SNPs positions are randomized in one case. 

In Class 1, we used \textit{bwa-mem} alignment tool for both positive and negative sets. We were able to align all positive sequences and ~93\% of negative sequences. While we cannot directly compare to the baseline given less negative sequences, we achieved similar ($F_1$ = 93.01) performance on this dataset compared to the baseline performance ($F_1$ = 93.5).

In Class 2, negative sequences are generated by randomly sampling SNPs from all 121 possible variants and imputing the sequences with them. This makes the task easier because the positive samples only covers 20.66\% of these variants while the negative samples contain SNPs from all 121 possible variants, leading to a high $F_1$ score (97.46).

In Class 3, negative sequences are generated by randomly inserting the same SNPs found in the positive sequences, preserving their distribution. 
This approach is possibly the most effective, as it maintains the distribution of SNPs while introducing variation. It outperforms the baseline, achieving a higher $F_1$ score (94.1) compared to the baseline (93.5) demonstrating the model's ability to benefit from variant-encoded sequences.

Class 4 presents the toughest challenge, where SNPs are randomly inserted into the positive sequences, creating negatives. Using a pretrained model ($F_1$=64.6) illustrates its ability to capture the natural SNP locations, to better understand and adapt to natural biological variations, when compared to the model trained from scratch ($F_1$=57.04), which struggles more with this disruption.

While the above results are promising, to draw meaningful conclusions, we need more datasets and further research to thoroughly evaluate the pretrained model. The methods for generating negative samples provide valuable insights, but a broader set of experiments is required to better assess how well the model captures natural SNP patterns and generalizes across different contexts.

%Hence the context of genetic variations can be learned by our foundation model through pre-training, which facilitates in certain cases the inference of SNP-related functional tasks. It is not clear why the lenti-MPRA task although beating DNA-BERT2, did not do better when the variants were included in the dataset. One hypothesis is that it is necessary to implement a different approach for fine-tuning or generating data that includes examples with measured effects of the variants in gene expression. Finally, it is of particular interest that BMFM-DNA can be used to generate embeddings of SNPs surrounded by their DNA regions to predict SNP-to-disease associations and perform comparable to SOTA models such as EVE \cite{frazer2021EVE}.

% Model Params
% Training Schemes (Loss: Binary Cross-Entropy Losses with Logits for multi-label classification; )
% Evaluation Results ()

% We separately trained two models, one for ClinVar and another for GWAS Catalog, by using pairs of SNP and Trait associations: i) ClinVar: Training Set (N = 511,397) and Validation Set (N = 219,171); ii) GWAS Catalog: Training Set (N = 187,595) and Validation Set (N = 80,398). 

%\subsubsection{Zero-shot tasks}

\section{Conclusion}
Through pre-training on 20 million genomic variants from the human genome, we developed \texttt{BMFM-DNA}, a unique SNP-aware foundation model that encodes both the standard DNA sequences and its natural variations enabling to capture the variant effects. Our foundation models trained using the human genome achieved similar predictive performance when compared with DNABERT-2, which was trained approximately three times longer on a much larger multi-species dataset. Our variant-encoded pretrained model, \texttt{BMFM-DNA-SNP}, performed consistently better across most tasks compared to the reference genome only model, \texttt{BMFM-DNA-REF}, demonstrating the importance of encoding the reference and sequence variations when building DNALMs. We also analyzed the impact of different variant-encoding strategies on the promoter detection task, demonstrating the model's ability to detect the context where natural biological variations occur.

%In particular this approach allows the detection of promoter regions with improved performance, while the prediction of the functional effects of nucleotide variation derived from regulatory regions remains challenging as demonstrated by the lenti-MPRA results. Therefore, we show for the first time that the information encoded in structural variants, such as SNPs, can be transferred during pre-training to DNALMs just based on an MLM approach. 

As we continue to expand the software package with additional capabilities, including multiomics data integration and novel modeling strategies, we invite the research community to collaborate in exploring optimal representations and strategies for building meaningful and state-of-the-art DNALMs. In particular, as we move forward to better understand the biological information stored in nucleotide variants that the model captured, we will benchmark using datasets with longer contextual sequences. Further, we also aim to extend the \texttt{BMFM-DNA} models to incorporate epigenomic information, enabling a more comprehensive representation of regulatory genomics.

%\bibliographystyle{sn-nature}
%\bibliography{main}% common bib file

\begin{thebibliography}{10}
\expandafter\ifx\csname url\endcsname\relax
  \def\url#1{\burl{#1}}\fi
\expandafter\ifx\csname urlprefix\endcsname\relax\def\urlprefix{URL }\fi
\providecommand{\bibinfo}[2]{#2}
\providecommand{\eprint}[2][]{\url{#2}}
\providecommand{\doi}[1]{\url{https://doi.org/#1}}
\bibcommenthead

\bibitem{international2001}
\bibinfo{author}{international consortium, T. H. G.~P.}
\newblock \bibinfo{title}{Initial sequencing and analysis of the human genome}.
\newblock \emph{\bibinfo{journal}{nature}} \textbf{\bibinfo{volume}{409}},
  \bibinfo{pages}{860--921} (\bibinfo{year}{2001}).

\bibitem{encode2012}
\bibinfo{author}{Consortium, E.~P.} \emph{et~al.}
\newblock \bibinfo{title}{An integrated encyclopedia of dna elements in the
  human genome}.
\newblock \emph{\bibinfo{journal}{Nature}} \textbf{\bibinfo{volume}{489}},
  \bibinfo{pages}{57} (\bibinfo{year}{2012}).

\bibitem{jaganathan2019}
\bibinfo{author}{Jaganathan, K.} \emph{et~al.}
\newblock \bibinfo{title}{Predicting splicing from primary sequence with deep
  learning}.
\newblock \emph{\bibinfo{journal}{Cell}} \textbf{\bibinfo{volume}{176}},
  \bibinfo{pages}{535--548} (\bibinfo{year}{2019}).

\bibitem{poplin2018}
\bibinfo{author}{Poplin, R.} \emph{et~al.}
\newblock \bibinfo{title}{A universal snp and small-indel variant caller using
  deep neural networks}.
\newblock \emph{\bibinfo{journal}{Nature biotechnology}}
  \textbf{\bibinfo{volume}{36}}, \bibinfo{pages}{983--987}
  (\bibinfo{year}{2018}).

\bibitem{zhou2015}
\bibinfo{author}{Zhou, J.} \& \bibinfo{author}{Troyanskaya, O.~G.}
\newblock \bibinfo{title}{Predicting effects of noncoding variants with deep
  learning--based sequence model}.
\newblock \emph{\bibinfo{journal}{Nature methods}}
  \textbf{\bibinfo{volume}{12}}, \bibinfo{pages}{931--934}
  (\bibinfo{year}{2015}).

\bibitem{avsec2021}
\bibinfo{author}{Avsec, {\v{Z}}.} \emph{et~al.}
\newblock \bibinfo{title}{Base-resolution models of transcription-factor
  binding reveal soft motif syntax}.
\newblock \emph{\bibinfo{journal}{Nature genetics}}
  \textbf{\bibinfo{volume}{53}}, \bibinfo{pages}{354--366}
  (\bibinfo{year}{2021}).

\bibitem{bommasani_opportunities_2022}
\bibinfo{author}{Bommasani, R.} \emph{et~al.}
\newblock \bibinfo{title}{On the {Opportunities} and {Risks} of {Foundation}
  {Models}} (\bibinfo{year}{2022}).
\newblock \urlprefix\url{http://arxiv.org/abs/2108.07258}.
\newblock \bibinfo{note}{ArXiv:2108.07258 [cs]}.

\bibitem{zhou_comprehensive_2024}
\bibinfo{author}{Zhou, C.} \emph{et~al.}
\newblock \bibinfo{title}{A comprehensive survey on pretrained foundation
  models: a history from {BERT} to {ChatGPT}}.
\newblock \emph{\bibinfo{journal}{International Journal of Machine Learning and
  Cybernetics}}  (\bibinfo{year}{2024}).
\newblock \urlprefix\url{https://link.springer.com/10.1007/s13042-024-02443-6}.

\bibitem{si_foundation_2024}
\bibinfo{author}{Si, Y.} \emph{et~al.}
\newblock \bibinfo{title}{Foundation models in molecular biology}.
\newblock \emph{\bibinfo{journal}{Biophysics Reports}}
  \textbf{\bibinfo{volume}{10}}, \bibinfo{pages}{135} (\bibinfo{year}{2024}).
\newblock \urlprefix\url{https://pmc.ncbi.nlm.nih.gov/articles/PMC11252241/}.

\bibitem{vaswani2017}
\bibinfo{author}{Vaswani, A.} \emph{et~al.}
\newblock \bibinfo{title}{Attention is all you need}.
\newblock \emph{\bibinfo{journal}{Advances in neural information processing
  systems}} \textbf{\bibinfo{volume}{30}} (\bibinfo{year}{2017}).

\bibitem{haig2004}
\bibinfo{author}{Haig, D.} \& \bibinfo{author}{Henikoff, S.}
\newblock \bibinfo{title}{Genomes and evolution: Deciphering the genomic
  palimpsest}.
\newblock \emph{\bibinfo{journal}{Current Opinion in Genetics \& Development}}
  \textbf{\bibinfo{volume}{14}}, \bibinfo{pages}{599--602}
  (\bibinfo{year}{2004}).

\bibitem{maurano2012}
\bibinfo{author}{Maurano, M.~T.} \emph{et~al.}
\newblock \bibinfo{title}{Systematic localization of common disease-associated
  variation in regulatory dna}.
\newblock \emph{\bibinfo{journal}{Science}} \textbf{\bibinfo{volume}{337}},
  \bibinfo{pages}{1190--1195} (\bibinfo{year}{2012}).

\bibitem{zhou2023dnabert}
\bibinfo{author}{Zhou, Z.} \emph{et~al.}
\newblock \bibinfo{title}{Dnabert-2: Efficient foundation model and benchmark
  for multi-species genome}.
\newblock \emph{\bibinfo{journal}{arXiv preprint arXiv:2306.15006}}
  (\bibinfo{year}{2023}).

\bibitem{fishman2025}
\bibinfo{author}{Fishman, V.} \emph{et~al.}
\newblock \bibinfo{title}{Gena-lm: a family of open-source foundational dna
  language models for long sequences}.
\newblock \emph{\bibinfo{journal}{Nucleic Acids Research}}
  \textbf{\bibinfo{volume}{53}}, \bibinfo{pages}{gkae1310}
  (\bibinfo{year}{2025}).

\bibitem{dalla2024nucleotide}
\bibinfo{author}{Dalla-Torre, H.} \emph{et~al.}
\newblock \bibinfo{title}{Nucleotide transformer: building and evaluating
  robust foundation models for human genomics}.
\newblock \emph{\bibinfo{journal}{Nature Methods}} \bibinfo{pages}{1--11}
  (\bibinfo{year}{2024}).

\bibitem{nguyen2024sequence}
\bibinfo{author}{Nguyen, E.} \emph{et~al.}
\newblock \bibinfo{title}{Sequence modeling and design from molecular to genome
  scale with evo}.
\newblock \emph{\bibinfo{journal}{Science}} \textbf{\bibinfo{volume}{386}},
  \bibinfo{pages}{eado9336} (\bibinfo{year}{2024}).

\bibitem{sanabria2024}
\bibinfo{author}{Sanabria, M.}, \bibinfo{author}{Hirsch, J.},
  \bibinfo{author}{Joubert, P.~M.} \& \bibinfo{author}{Poetsch, A.~R.}
\newblock \bibinfo{title}{Dna language model grover learns sequence context in
  the human genome}.
\newblock \emph{\bibinfo{journal}{Nature Machine Intelligence}}
  \textbf{\bibinfo{volume}{6}}, \bibinfo{pages}{911--923}
  (\bibinfo{year}{2024}).

\bibitem{chen2022}
\bibinfo{author}{Chen, K.~M.}, \bibinfo{author}{Wong, A.~K.},
  \bibinfo{author}{Troyanskaya, O.~G.} \& \bibinfo{author}{Zhou, J.}
\newblock \bibinfo{title}{A sequence-based global map of regulatory activity
  for deciphering human genetics}.
\newblock \emph{\bibinfo{journal}{Nature genetics}}
  \textbf{\bibinfo{volume}{54}}, \bibinfo{pages}{940--949}
  (\bibinfo{year}{2022}).

\bibitem{tang2024evaluating}
\bibinfo{author}{Tang, Z.}, \bibinfo{author}{Somia, N.}, \bibinfo{author}{Yu,
  Y.} \& \bibinfo{author}{Koo, P.~K.}
\newblock \bibinfo{title}{Evaluating the representational power of pre-trained
  dna language models for regulatory genomics}.
\newblock \emph{\bibinfo{journal}{bioRxiv}}  (\bibinfo{year}{2024}).

\bibitem{patel2024dart}
\bibinfo{author}{Patel, A.} \emph{et~al.}
\newblock \bibinfo{title}{Dart-eval: A comprehensive dna language model
  evaluation benchmark on regulatory dna}.
\newblock \emph{\bibinfo{journal}{arXiv preprint arXiv:2412.05430}}
  (\bibinfo{year}{2024}).

\bibitem{agarwal2025}
\bibinfo{author}{Agarwal, V.} \emph{et~al.}
\newblock \bibinfo{title}{Massively parallel characterization of
  transcriptional regulatory elements}.
\newblock \emph{\bibinfo{journal}{Nature}} \bibinfo{pages}{1--10}
  (\bibinfo{year}{2025}).

\bibitem{schneider2017}
\bibinfo{author}{Schneider, V.~A.} \emph{et~al.}
\newblock \bibinfo{title}{Evaluation of grch38 and de novo haploid genome
  assemblies demonstrates the enduring quality of the reference assembly}.
\newblock \emph{\bibinfo{journal}{Genome research}}
  \textbf{\bibinfo{volume}{27}}, \bibinfo{pages}{849--864}
  (\bibinfo{year}{2017}).

\bibitem{smigielski2000}
\bibinfo{author}{Smigielski, E.~M.}, \bibinfo{author}{Sirotkin, K.},
  \bibinfo{author}{Ward, M.} \& \bibinfo{author}{Sherry, S.~T.}
\newblock \bibinfo{title}{dbsnp: a database of single nucleotide
  polymorphisms}.
\newblock \emph{\bibinfo{journal}{Nucleic acids research}}
  \textbf{\bibinfo{volume}{28}}, \bibinfo{pages}{352--355}
  (\bibinfo{year}{2000}).

\bibitem{dreos2013epdnew}
\bibinfo{author}{Dreos, R.}, \bibinfo{author}{Ambrosini, G.},
  \bibinfo{author}{Cavin~P{\'e}rier, R.} \& \bibinfo{author}{Bucher, P.}
\newblock \bibinfo{title}{Epd and epdnew, high-quality promoter resources in
  the next-generation sequencing era}.
\newblock \emph{\bibinfo{journal}{Nucleic acids research}}
  \textbf{\bibinfo{volume}{41}}, \bibinfo{pages}{D157--D164}
  (\bibinfo{year}{2013}).

\bibitem{oubounyt2019deepromoter}
\bibinfo{author}{Oubounyt, M.}, \bibinfo{author}{Louadi, Z.},
  \bibinfo{author}{Tayara, H.} \& \bibinfo{author}{Chong, K.~T.}
\newblock \bibinfo{title}{Deepromoter: robust promoter predictor using deep
  learning}.
\newblock \emph{\bibinfo{journal}{Frontiers in genetics}}
  \textbf{\bibinfo{volume}{10}}, \bibinfo{pages}{286} (\bibinfo{year}{2019}).

\bibitem{wang2019}
\bibinfo{author}{Wang, R.}, \bibinfo{author}{Wang, Z.}, \bibinfo{author}{Wang,
  J.} \& \bibinfo{author}{Li, S.}
\newblock \bibinfo{title}{Splicefinder: ab initio prediction of splice sites
  using convolutional neural network}.
\newblock \emph{\bibinfo{journal}{BMC bioinformatics}}
  \textbf{\bibinfo{volume}{20}}, \bibinfo{pages}{1--13} (\bibinfo{year}{2019}).

\bibitem{pinero2020disgenet}
\bibinfo{author}{Pi{\~n}ero, J.} \emph{et~al.}
\newblock \bibinfo{title}{The disgenet knowledge platform for disease genomics:
  2019 update}.
\newblock \emph{\bibinfo{journal}{Nucleic acids research}}
  \textbf{\bibinfo{volume}{48}}, \bibinfo{pages}{D845--D855}
  (\bibinfo{year}{2020}).

\bibitem{cerezo2025}
\bibinfo{author}{Cerezo, M.} \emph{et~al.}
\newblock \bibinfo{title}{The nhgri-ebi gwas catalog: standards for
  reusability, sustainability and diversity}.
\newblock \emph{\bibinfo{journal}{Nucleic acids research}}
  \textbf{\bibinfo{volume}{53}}, \bibinfo{pages}{D998--D1005}
  (\bibinfo{year}{2025}).

\bibitem{landrum2019clinvar}
\bibinfo{author}{Landrum, M.~J.} \emph{et~al.}
\newblock \bibinfo{title}{Clinvar: improvements to accessing data}.
\newblock \emph{\bibinfo{journal}{Nucleic Acids Research}}
  \textbf{\bibinfo{volume}{48}}, \bibinfo{pages}{D835--D844}
  (\bibinfo{year}{2019}).

\bibitem{malone2010}
\bibinfo{author}{Malone, J.} \emph{et~al.}
\newblock \bibinfo{title}{Modeling sample variables with an experimental factor
  ontology}.
\newblock \emph{\bibinfo{journal}{Bioinformatics}}
  \textbf{\bibinfo{volume}{26}}, \bibinfo{pages}{1112--1118}
  (\bibinfo{year}{2010}).

\bibitem{vasilevsky2022}
\bibinfo{author}{Vasilevsky, N.~A.} \emph{et~al.}
\newblock \bibinfo{title}{Mondo: unifying diseases for the world, by the
  world}.
\newblock \emph{\bibinfo{journal}{MedRxiv}} \bibinfo{pages}{2022--04}
  (\bibinfo{year}{2022}).

\bibitem{li2013aligning}
\bibinfo{author}{Li, H.}
\newblock \bibinfo{title}{Aligning sequence reads, clone sequences and assembly
  contigs with bwa-mem}.
\newblock \emph{\bibinfo{journal}{arXiv preprint arXiv:1303.3997}}
  (\bibinfo{year}{2013}).

\bibitem{vaswani2017attention}
\bibinfo{author}{Vaswani, A.}
\newblock \bibinfo{title}{Attention is all you need}.
\newblock \emph{\bibinfo{journal}{Advances in Neural Information Processing
  Systems}}  (\bibinfo{year}{2017}).

\bibitem{warner2024}
\bibinfo{author}{Warner, B.} \emph{et~al.}
\newblock \bibinfo{title}{Smarter, better, faster, longer: A modern
  bidirectional encoder for fast, memory efficient, and long context finetuning
  and inference}.
\newblock \emph{\bibinfo{journal}{arXiv preprint arXiv:2412.13663}}
  (\bibinfo{year}{2024}).

\bibitem{cahyawijaya2022}
\bibinfo{author}{Cahyawijaya, S.} \emph{et~al.}
\newblock \bibinfo{title}{Snp2vec: scalable self-supervised pre-training for
  genome-wide association study}.
\newblock \emph{\bibinfo{journal}{arXiv preprint arXiv:2204.06699}}
  (\bibinfo{year}{2022}).

\bibitem{sennrich2015}
\bibinfo{author}{Sennrich, R.}, \bibinfo{author}{Haddow, B.} \&
  \bibinfo{author}{Birch, A.}
\newblock \bibinfo{title}{Neural machine translation of rare words with subword
  units}.
\newblock \emph{\bibinfo{journal}{arXiv preprint arXiv:1508.07909}}
  (\bibinfo{year}{2015}).

\bibitem{kudo2018}
\bibinfo{author}{Kudo, T.} \& \bibinfo{author}{Richardson, J.}
\newblock \bibinfo{title}{Sentencepiece: A simple and language independent
  subword tokenizer and detokenizer for neural text processing}.
\newblock \emph{\bibinfo{journal}{arXiv preprint arXiv:1808.06226}}
  (\bibinfo{year}{2018}).

\bibitem{litdata2023}
\bibinfo{author}{Chaton, T.} \& \bibinfo{author}{AI, L.}
\newblock \bibinfo{title}{Litdata: Transform datasets at scale. optimize
  datasets for fast ai model training.}
\newblock \bibinfo{howpublished}{\url{https://github.com/Lightning-AI/litdata}}
  (\bibinfo{year}{2023}).
\newblock \bibinfo{note}{Accessed: 2025-04-09}.

\bibitem{devlin2018bert}
\bibinfo{author}{Devlin, J.}
\newblock \bibinfo{title}{Bert: Pre-training of deep bidirectional transformers
  for language understanding}.
\newblock \emph{\bibinfo{journal}{arXiv preprint arXiv:1810.04805}}
  (\bibinfo{year}{2018}).

\bibitem{frazer2021EVE}
\bibinfo{author}{Frazer, J.} \emph{et~al.}
\newblock \bibinfo{title}{Disease variant prediction with deep generative
  models of evolutionary data}.
\newblock \emph{\bibinfo{journal}{Nature}} \textbf{\bibinfo{volume}{599}},
  \bibinfo{pages}{91--95} (\bibinfo{year}{2021}).

\bibitem{rafi2024}
\bibinfo{author}{Rafi, A.~M.} \emph{et~al.}
\newblock \bibinfo{title}{A community effort to optimize sequence-based deep
  learning models of gene regulation}.
\newblock \emph{\bibinfo{journal}{Nature biotechnology}} \bibinfo{pages}{1--11}
  (\bibinfo{year}{2024}).

\bibitem{cheng2023}
\bibinfo{author}{Cheng, J.} \emph{et~al.}
\newblock \bibinfo{title}{Accurate proteome-wide missense variant effect
  prediction with alphamissense}.
\newblock \emph{\bibinfo{journal}{Science}} \textbf{\bibinfo{volume}{381}},
  \bibinfo{pages}{eadg7492} (\bibinfo{year}{2023}).

\end{thebibliography}

\end{document}